\documentclass[aps,prl,twocolumn,showpacs,floatfix]{revtex4}

\usepackage{graphicx}
\usepackage{dcolumn}
\usepackage{bm}
\usepackage{hyperref}

\usepackage{amsmath,amsfonts,amssymb,amsthm}
\usepackage{braket}
\usepackage{slashed}
\usepackage{comment}
\usepackage{float}
\usepackage{mhchem}
\usepackage{tabularx}
\usepackage{times}
\usepackage{nicefrac}
\usepackage{epsf}
\usepackage{bm}
\usepackage{bbm}
\usepackage{color}

\newcommand{\dgint}{\Delta g_\text{int}}
\newcommand{\dgqed}{\Delta g_\text{QED}}

\newcommand{\dgnuc}{\Delta g_\text{nuc}}
\usepackage{siunitx}

\usepackage{verbatim}

\usepackage{pstricks,pst-node,pst-text,pst-3d}

\usepackage[normalem]{ulem}  
\newrgbcolor{cLemon}{0.9 0.4 0.5}
\newrgbcolor{cMarine}{0.1 0.3 0.7}
\newrgbcolor{cDeepBlue}{0.05 0.1 0.5}
\newrgbcolor{cLightBlue}{0.38 0.55 0.7}
\newrgbcolor{cGreen}{0.0 0.6 0.0}
\newrgbcolor{cGreenn}{0.0 0.7 0.0}
\newrgbcolor{cGreennn}{0.0 0.8 0.0}
\newrgbcolor{cGreennnn}{0.0 0.9 0.0}
\newrgbcolor{cDarkGreen}{0.3 0.7 0.3}
\newrgbcolor{cUV}{0.6 0.0 0.6}
\newrgbcolor{cIR}{0.6 0.0 0.0}
\newrgbcolor{malina}{0.8 0.2 0.4}
\newrgbcolor{Lemon}{1.0 0.2 0.5}
\newrgbcolor{electron}{0.1 0.2 0.5}
\newrgbcolor{out}{0.05 0.1 0.3}
\newrgbcolor{LightBlue}{0.68 0.85 0.9}
\newrgbcolor{cRed}{0.6 0.0 0.0}
\newrgbcolor{cRedd}{0.7 0.0 0.0}
\newrgbcolor{cReddd}{0.8 0.0 0.0}
\newrgbcolor{cRedddd}{0.9 0.0 0.0}
\newrgbcolor{cOrange}{0.8 0.4 0.0}
\newrgbcolor{cOrangge}{0.6 0.3 0.0}
\newrgbcolor{cYellow}{0.5 0.8 0.4}
\newrgbcolor{cGrey}{0.5 0.5 0.5}
\newcommand{\be}{\begin{eqnarray}}
\newcommand{\ee}{\end{eqnarray}}

\newcommand{\aZ}{\alpha Z}
\preprint{APS/123-QED}
\begin{document}
\thispagestyle{empty}
\title{
\emph{g} Factor of Lithiumlike Silicon and Calcium:\\ Resolving the Disagreement between Theory and Experiment
}
\author{V. P. Kosheleva,$^{1,2,3}$ A. V. Volotka,$^{4}$ D. A. Glazov,$^{5}$ D. V. Zinenko,$^{5}$ and S. Fritzsche$^{1,2,3}$}

\affiliation{
$^1$ Theoretisch-Physikalisches Institut, Friedrich-Schiller-Universität Jena, Max-Wien-Platz 1, 07743 Jena, Germany\\
$^2$ Helmholtz-Institut Jena, Fr\"obelstieg 3, 07743 Jena, Germany\\
$^3$ GSI Helmholtzzentrum für Schwerionenforschung GmbH, Planckstraße 1, 64291 Darmstadt, Germany\\
$^4$ School of Physics and Engineering, ITMO University, Kronverkskiy pr. 49, 197101 St. Petersburg, Russia\\
$^5$ Department of Physics, St. Petersburg State University, Universitetskaya nab. 7/9, 199034 St. Petersburg, Russia\\
}
%
\begin{abstract}
The bound-electron \emph{g} factor is a stringent tool for tests of the Standard Model and the search for new physics.
The comparison between an experiment on the \emph{g} factor of lithiumlike silicon and the two recent theoretical values revealed the discrepancies of $1.7\sigma$ [D. A. Glazov \textit{et al}., Phys. Rev. Lett. \textbf{123}, 173001 (2019)] and $5.2\sigma$ [V. A. Yerokhin \textit{et al}., Phys. Rev. A \textbf{102}, 022815 (2020)].
To identify the reason for this disagreement, we accomplish large-scale high-precision computation of the interelectronic-interaction and many-electron QED corrections.
The calculations are performed within the extended Furry picture of QED, and the dependence of the final values on the choice of the binding potential is carefully analyzed.
As a result, we significantly improve the agreement between the theory and experiment for the \emph{g} factor of lithiumlike silicon. We also report the most accurate theoretical prediction to date for lithiumlike calcium, which perfectly agrees with the experimental value.

\end{abstract}
%
\maketitle
%
%
%
%
%
%
%
%
\textit{Introduction.}
Over the past decades, the Zeeman effect in highly charged ions has been a subject of intense theoretical and experimental investigations.
Nowadays, the bound-electron \emph{g} factor is measured with a relative accuracy of a few parts in $10^{11}$ in H-like carbon and silicon ions \cite{sturm:2011:023002, sturm:2013:R030501, sturm:2014:467}. 
%
%
%
These measurements, accompanied by impressive theoretical studies  \cite{blundell:1997:1857, persson:1997:R2499, beier:2000:79, karshenboim:2000:380, karshenboim:2001:81, glazov:2002:408, shabaev:2002:091801, nefiodov:2002:081802, yerokhin:2002:143001, yerokhin:2004:052503, pachucki:2004:150401, pachucki:2005:022108, lee:2005:052501, jentschura:2009:044501, yerokhin:2013:042502, czarnecki:2016:060501, yerokhin:2017:060501, czarnecki:2018:043203, czarnecki:2020:050801}, have led to the most accurate up-to-date value of the electron mass \cite{sturm:2014:467, koehler:2015:144032, zatorski:2017:012502}.
%
%
Furthermore, the present experimental techniques also enable high-precision \emph{g}-factor measurements in few-electron ions \cite{wagner:2013:033003, lindenfels:2013:023412, koehler:2016:10246, glazov:2019:173001, arapoglou:2019:253001}.
In particular, in the recent experiments for Li-like ions, the results with eleven significant digits are reported~\cite{koehler:2016:10246, glazov:2019:173001}, thus reaching an accuracy comparable to that for H-like ions.
\\ 
\indent
The unprecedented precision achieved both in experiments and in theory brings the bound-electron \emph{g} factor to the category of observables that define our understanding of fundamental physics.
For example, the measurement of the \emph{g}-factor isotope shift with Li-like calcium ions \cite{koehler:2016:10246} has opened a possibility to test the relativistic nuclear recoil theory in the presence of magnetic field and paved the way to probe bound-state QED effects beyond the Furry picture in the strong-field regime \cite{shabaev:2017:263001, malyshev:2017:731}.
The high-precision bound-electron \emph{g}-factor experiments combined with theoretical studies are expected to provide an independent determination of the fine structure constant $\alpha$ \cite{shabaev:2006:253002, volotka:2014:023002, yerokhin:2016:022502}.
Moreover, one can search for the effects beyond the Standard Model~\cite{debierre:2020:135527}.
While calcium is the heaviest system measured to date, the most interesting effects, including new physics, generally grow with the nuclear charge number $Z$. Thus, the middle-$Z$ ions primarily serve as a prototype to verify the theoretical methods, which still need further development to match the experimental precision and to realize these fascinating ideas eventually.
\\ 
\indent
The first Penning-trap \emph{g}-factor measurements with Li-like ions were performed for silicon \cite{wagner:2013:033003} and calcium \cite{koehler:2016:10246} with an uncertainty of about $10^{-9}$.
Recently, the 15-fold improved experimental value for $^{28}$Si$^{11+}$ was published~\cite{glazov:2019:173001}: currently, it is the most accurate \emph{g}-factor value for the few-electron ions.
To match the experimental precision, a multitude of the QED and nuclear effects should be rigorously taken into account in theory.
The major difficulty of Li-like systems in comparison to the H-like ones consists in the many-electron contributions.
The recent progress in the many-electron QED calculations includes the evaluation of the screened QED diagrams~\cite{volotka:2009:033005, glazov:2010:062112, andreev:2012:022510, volotka:2014:253004, yerokhin:2020:022815}, the two-photon-exchange diagrams~\cite{volotka:2012:073001, wagner:2013:033003, volotka:2014:253004, yerokhin:2021:022814}, and various higher-order effects \cite{yerokhin:2017:062511, glazov:2019:173001, cakir:2020:062513}. 
The nuclear recoil effect in Li-like ions was addressed recently in Refs.~\cite{shabaev:2017:263001, malyshev:2017:731, shabaev:2018:032512}.
In Ref.~\cite{glazov:2019:173001}, the most accurate at that time theoretical \emph{g}-factor value for $^{28}$Si$^{11+}$ was obtained, $g_\text{th,2019} = 2.000\,889\,894\,4\,(34)$, and it was found to be $1.7 \sigma$ away from the experimental value, $g_\text{exp,2019} = 2.000\,889\,888\,45\,(14)$, presented ibid.
The improvement was achieved mainly due to the accurate treatment of the higher-order effects within the perturbation theory. The calculations were performed within the extended Furry picture, where the Dirac equation includes an effective screening potential.
\\ 
\indent
In attempt to resolve this discrepancy, an independent evaluation of the screened QED diagrams was undertaken by Yerokhin {\it et al.} \cite{yerokhin:2020:022815}.
In contrast to Ref.~\cite{glazov:2019:173001}, the calculations were performed within the standard Furry picture, i.e., based on the Dirac equation with the Coulomb potential.
The higher-order effects were evaluated within the nonrelativistic quantum electrodynamics (NRQED) approach to the leading order in $\alpha Z$.
As a result, a new theoretical value for $^{28}$Si$^{11+}$ was obtained: $ g_\text{th,2020} =  2.000\,889\,896\,3\,(15)$~\cite{yerokhin:2020:022815}.
For both $g_\text{th,2019}$ and $g_\text{th,2020}$, the theoretical error bar is determined by the numerical uncertainty of the calculated contributions and by estimating the unknown higher-order many-electron QED effects.
Here, we note that even though in the case of the original Furry picture the higher-order effects are generally more significant than those in the extended Furry picture, the corresponding uncertainty suggested in Ref.~\cite{yerokhin:2020:022815} is twice smaller than in Ref.~\cite{glazov:2019:173001}.
Overall, $g_\text{th,2019}$ and $g_\text{th,2020}$ are in fair agreement within the quoted error bars.
However, the theoretical value from Ref.~\cite{yerokhin:2020:022815} disagrees by about $5.2\sigma$ with experiment and, therefore, the \emph{g}-factor ``puzzle'' has only got worse. 
Just recently, Yerokhin {\it et al.} accomplished an independent evaluation of the two-photon-exchange contribution~\cite{yerokhin:2021:022814} and presented new results for Li-like silicon and calcium. For silicon, the $3.1\sigma$ disagreement remains, somewhat smaller than before~\cite{yerokhin:2020:022815}. The calcium \emph{g}-factor value differs by $4.2\sigma$ from the experimental value~\cite{koehler:2016:10246}.
\\ 
\indent
%
%
The screened QED and interelectronic-interaction effects represent the main challenge for theory. So far, rigorous evaluation of these contributions has been accomplished only by two groups.
In this Letter, we further scrutinize the many-electron QED effects to shed light on the persisting discrepancy.
We have performed large-scale QED calculations in the extended Furry picture for different screening potentials and found that our present \emph{g}-factor values are in fair agreement with experiment for both silicon and calcium. 
\\ 
\indent
%
%
\textit{Basic theory.} 
The ground-state \emph{g} factor of lithiumlike ion with a spinless nucleus can be written as
\be
\label{eq:gtotal}
  g = g^{(0)}_\mathrm{C} + \dgint + \dgqed + \dgnuc\,.
\ee
Here, $g^{(0)}_\mathrm{C}$ is the lowest-order \emph{g}-factor value obtained for Coulomb potential, $\dgint$ is the interelectronic-interaction correction, $\dgqed$ is the QED correction, and $\dgnuc$ stands for the nuclear recoil and nuclear polarization effects. 
Below, we focus on $\dgint$ and $\dgqed$. Each of these contributions can be expanded within the bound-state QED perturbation theory,
\be
  \Delta g =\Delta g^{(1)} + \Delta g^{(2)} + \Delta g^{(3+)} \,,
\label{eq:PT}
\ee
where the superscript $i$ refers to the $i$th order in $\alpha$, and $\Delta g^{(3+)}$ includes all the higher orders.
So far, only three terms have been rigorously evaluated, i.e., to all orders in $\alpha Z$ without any further approximations: $\Delta g^{(1)}_\text{int}$ --- one-photon exchange, $\Delta g^{(2)}_\text{int}$ --- two-photon exchange, and $\Delta g^{(1)}_\text{QED}$ --- one-electron self-energy and vacuum polarization.
The second-order QED correction $\Delta g^{(2)}_\text{QED}$ can be split into two parts: one-electron two-loop QED term $\Delta g^{(2)}_\text{QED-1e}$ and many-electron (screened) QED term $\Delta g^{(2)}_\text{QED-me}$.
While the evaluation of the former is still in progress \cite{yerokhin:2013:042502, sikora:2020:012002, debierre:2021:030802}, the latter was independently computed in Refs.~\cite{volotka:2009:033005, glazov:2010:062112} and \cite{yerokhin:2020:022815}.
The terms that are not yet known to all orders in $\alpha Z$ are evaluated approximately, e.g., within the $\alpha Z$ expansion or by employing some effective operators.
We present these terms in the following form:
\be
\Delta g^{(i)} = \Delta g^{(i)}_\text{L} + \Delta g^{(i)}_\text{H}\,,
\ee
where $\Delta g^{(i)}_\text{L}$ denotes the leading-order part that is taken into account, and $\Delta g^{(i)}_\text{H}$ is the presently unknown higher-order part whose value needs estimation to ascribe the uncertainty to $\Delta g^{(i)}$.
Note that the leading-order terms $\Delta g^{(i)}_\text{L}$ can be defined in different ways depending on the calculation method, e.g., NRQED or the Breit approximation.
In general, the higher-order part $\Delta g^{(i)}_\text{H}$ is suppressed by the factor $(\alpha Z)^2$ in comparison to $\Delta g^{(i)}_\text{L}$.
%
%
%
%
Since the rigorous treatment of $\Delta g^{(i)}$ is presently limited by the second order, the higher-order terms are evaluated approximately, i.e., including only the $\Delta g^{(3+)}_\text{L}$ part.  
This can be accomplished via NRQED approach \cite{yerokhin:2017:062511}, by employing the configuration interaction method \cite{bratsev:1977:2655,glazov:2004:062104}, or within the recursive perturbation theory \cite{glazov:2017:46,glazov:2019:173001} in the Breit approximation.
Presently, the theoretical accuracy is mainly limited by the missing higher-order contributions $\Delta g^{(3+)}_\text{H}$, which can be estimated in several ways.
The first option is to use the higher-order term from the previous order of the perturbation theory: $(i)\, \Delta g^{(3+)}_\text{H} \simeq \Delta g^{(2)}_\text{H} / Z$. 
The second option, in contrary, is based purely on the value of the leading-order term from the same order of perturbation theory: $(ii)\, \Delta g^{(3+)}_\text{H} \simeq \Delta g^{(3+)}_\text{L} (\aZ)^2$.
Finally, the third method combines both of these schemes: $(iii)\, \Delta g^{(3+)}_\text{H} \simeq \Delta g^{(2)}_\text{H} ( \Delta g^{(3+)}_\text{L} / \Delta g^{(2)}_\text{L} )$.
%
%
\\
\indent
%
%
%
The higher-order contribution $g^{(3+)}$, including the presently unknown part $\Delta g^{(3+)}_\text{H}$, can be significantly reduced by introducing an effective local screening potential in the Dirac equation, the so-called extended Furry picture~\cite{sapirstein:2001:022502}.
The perturbation series is rearranged so that the dominant part of each order is transferred to the lower orders.
While each order individually depends on the choice of the screening potential, the total result should be potential-independent.
Since the higher-order part $\Delta g^{(3+)}_\text{H}$ is missing at present, the difference between the total values obtained with different screening potentials allows estimating the magnitude of $\Delta g^{(3+)}_\text{H}$.
%
%
As in our previous works, we employ the following screening potentials: core-Hartree (CH), Kohn-Sham (KS), Dirac-Hartree (DH), and Dirac-Slater (DS) one~\cite{sapirstein:2001:022502, glazov:2006:330}.
\\
\indent
%
%
\textit{Interelectronic interaction.}
The first-order correction $\Delta g^{(1)}_\text{int}$ is represented by the one-photon-exchange diagrams, and its calculation is rather straightforward \cite{shabaev:2002:062104}.
The next-order contribution $\Delta g^{(2)}_\text{int}$, developed previously in Refs.~\cite{volotka:2012:073001, volotka:2014:253004, kosheleva:2020:013364}, corresponds to the two-photon-exchange diagrams.
%
%
%
%
%
%
%
%
Finally, $\Delta g^{(3+)}_\text{int,L}$ is calculated by the recursive perturbation theory \cite{glazov:2017:46, glazov:2019:173001, kosheleva:2020:013364} within the Breit approximation.
For computational details on $\Delta g^{(2)}_\text{int}$ and $\Delta g^{(3+)}_\text{int,L}$, see Supplementary Material.
\\
\indent
%
%
%
\begin{table*}
\caption{\label{T2_var_pot} 
Interelectronic-interaction contributions $\dgint$ to the ground-state \emph{g} factor of Li-like silicon and calcium for different potentials: Coulomb, core-Hartree (CH), Kohn-Sham (KS), Dirac-Hartree (DH), and Dirac-Slater (DS), in units of $10^{-6}$.}  
\begin{center}
\setlength{\tabcolsep}{5pt}
\begin{tabular}{l S S S S S} \hline\hline
&\multicolumn{1}{c}{Coulomb} & \multicolumn{1}{c}{CH} & \multicolumn{1}{c}{KS} & \multicolumn{1}{c}{DH} & \multicolumn{1}{c}{DS} \\ \hline
\multicolumn{6}{l}{$Z = 14$}\\
$g^{(0)} - g^{(0)}_\mathrm{C}$ 
                    &               &348.2661      &341.3682      &353.1638      &329.1102      \\
$\dgint^{(1)}$      &321.5903       &-33.5491      &-25.0951      &-39.2815      &-11.7598      \\
$\dgint^{(2)}$      & -6.8782(1)    &  0.1362(1)   & -1.4838(1)   &  1.1237(1)   & -2.5910(1)   \\
                    & -6.8787(1)$^a$&  0.137$^b$ \\
$\Delta g^{(3+)}_\text{int,L}$
                    &  0.0934(21)   & -0.0443(10)  &  0.0202(12)  & -0.1952(18)  &  0.0505(12)  \\                    
                    &  0.0942(4)$^c$& -0.046(6)$^b$ \\
$\Delta g^{(3+)}_\text{int,H}$
                    &  0.0000(74)   &  0.0000(14)  &  0.0000(18)  &  0.0000(12)  &  0.0000(20)  \\
                    &  0.0000(14)$^a$ \\
Total               &314.8055(77)   &314.8089(17)  &314.8095(22)  &314.8107(22)  &314.8099(23)  \\                     
                    &314.8058(15)$^a$ \\[1mm]
\hline
\multicolumn{6}{l}{$Z = 20$}\\
$g^{(0)} - g^{(0)}_\mathrm{C}$ 
                    &               &505.2339      &494.1961      &513.4290      &475.2654      \\
$\dgint^{(1)}$      &461.1479       &-51.0429      &-38.3914      &-60.1565      &-18.3166      \\
$\dgint^{(2)}$      & -6.9338(1)    &  0.1291(1)   & -1.5297(1)   &  1.1550(1)   & -2.6958(1)   \\
                    & -6.9341(3)$^a$&  0.129$^b$ \\
$\Delta g^{(3+)}_\text{int,L}$
                    &  0.0661(17)   & -0.0300(8)   &  0.0155(12)  & -0.1359(13)  &  0.0388(13)  \\                    
                    &  0.0695(12)$^c$ \\
$\Delta g^{(3+)}_\text{int,H}$
                    &  0.0000(108)  &  0.0000(24)  &  0.0000(20)  &  0.0000(20)  &  0.0000(30)  \\
                    &  0.0000(22)$^a$ \\
Total               &454.2802(109)  &454.2902(25)  &454.2905(24)  &454.2915(24)  &454.2918(33)  \\
                    &454.2834(25)$^a$ \\
\hline\hline
\end{tabular}\\
$^a$ Yerokhin {\it et al.} (2021) \cite{yerokhin:2021:022814};
$^b$ Volotka {\it et al.} (2014) \cite{volotka:2014:253004};
$^c$ Yerokhin {\it et al.} (2017) \cite{yerokhin:2017:062511}.
\end{center}
\end{table*}
Table \ref{T2_var_pot} presents the results obtained for the interelectronic-interaction correction for Si$^{11+}$ and Ca$^{17+}$ ions. The zeroth-order value obtained in screening potential minus the Coulomb value, $g^{(0)} - g^{(0)}_\mathrm{C}$, is an important contribution to $\dgint$ in the extended Furry picture. The results for the two-photon exchange are compared to the corresponding results from Refs.~\cite{volotka:2014:253004, yerokhin:2021:022814}. As one can see, our values are one order of magnitude more accurate than those of Ref.~\cite{volotka:2014:253004}, while for the Coulomb potential, the marginal agreement is found with Ref.~\cite{yerokhin:2021:022814}. Our result for the third- and higher-order correction $\Delta g^{(3+)}_\text{int,L}$ obtained within the Breit approximation agrees well with the results of Refs.~\cite{yerokhin:2017:062511,yerokhin:2021:022814} obtained within the NRQED approach for the Coulomb potential.
\\
\indent
Before proceeding to the total results, we consider the uncertainty of the presently unknown higher-order term $\Delta g^{(3+)}_\text{int,H}$. For the silicon ion and the Coulomb potential, the three above-mentioned methods for its estimation yield $(i)$ 0.0037, $(ii)$ 0.0013, and $(iii)$ 0.0008 in units of $10^{-6}$. To select the appropriate method, we adopt the following reasoning. Once $\dgint$ is calculated rigorously, the total results should be the same for any binding potential. Hence, the present deviations between the Coulomb, CH, KS, DH, and DS results are due to $\Delta g^{(3+)}_\text{int,H}$. Thus, we choose the first (largest) uncertainty multiplied by a factor of 2 to provide the overlapping of the results. At the same time, the authors of Refs.~\cite{yerokhin:2017:062511, yerokhin:2021:022814} used the smallest uncertainty (third choice), multiplied by 1.5~\cite{yerokhin:2017:062511} and by 2~\cite{yerokhin:2021:022814}. As one can see from the Table, their total Coulomb result does not overlap with our values presented for other potentials, both for silicon and calcium.
\\
\indent
Finally, we average our total values over four screening potentials and obtain the interelectronic-interaction correction $314.8098(22)\times 10^{-6}$ for silicon and $454.2910(24)\times 10^{-6}$ for calcium.

%
%
\textit{QED corrections.}
\begin{table*}
\caption{\label{T3_var_pot}
QED corrections $\dgqed$ to the ground-state \emph{g} factor of Li-like silicon and calcium for different potentials: Coulomb, core-Hartree (CH), Kohn-Sham (KS), Dirac-Hartree (DH), and Dirac-Slater (DS), in units of $10^{-6}$. In the case of the Coulomb potential values are taken from Refs.~\cite{yerokhin:2020:022815, yerokhin:2021:022814}.}
\begin{center}
\begin{tabular}{l S S S S S}
\hline
\hline
&\multicolumn{1}{c}{Coulomb$^a$} & \multicolumn{1}{c}{CH} & \multicolumn{1}{c}{KS} & \multicolumn{1}{c}{DH} & \multicolumn{1}{c}{DS} \\ \hline
\multicolumn{6}{l}{$Z = 14$}\\
$\Delta g^{(1)}_\text{QED}$
              &2324.0439         &2323.8100         &2323.8106         &2323.8089         &2323.8227     \\
$\Delta g^{(2)}_\text{QED-1e}$
              &  -3.5463         &  -3.5460         &  -3.5460         &  -3.5460         &  -3.5460     \\
$\Delta g^{(2)}_\text{QED-me}$
              &  -0.2460(6)      &  -0.0074(17)     &  -0.0087(17)     &  -0.0064(16)     &  -0.0216(20) \\
$\Delta g^{(3+)}_\text{QED-1e}$
              &   0.0295         &   0.0295         &   0.0295         &   0.0295         &   0.0295     \\
$\Delta g^{(3+)}_\text{QED-me,L}$
              &   0.0099         &  -0.0003         &   0.0003         &  -0.0004         &   0.0009     \\
$\Delta g^{(3+)}_\text{QED-me,H}$
              &   0.0000(6)      &   0.0000         &   0.0000(2)      &   0.0000(1)      &   0.0000(1)  \\
Total         &2320.2910(8)      &2320.2858(17)     &2320.2857(17)     &2320.2856(16)     &2320.2855(20) \\[1mm]
\hline
\multicolumn{6}{l}{$Z = 20$}\\
$\Delta g^{(1)}_\text{QED}$
              &2325.5544         &2325.2019         &2325.1985         &2325.2025         &2325.2211     \\
$\Delta g^{(2)}_\text{QED-1e}$
              &  -3.5490(3)      &  -3.5484(3)      &  -3.5484(3)      &  -3.5484(3)      &  -3.5485(3)  \\
$\Delta g^{(2)}_\text{QED-me}$
              &  -0.3675(6)      &  -0.0220(17)     &  -0.0199(17)     &  -0.0228(15)     &  -0.0438(20) \\
$\Delta g^{(3+)}_\text{QED-1e}$
              &   0.0295         &   0.0295         &   0.0295         &   0.0295         &   0.0295     \\
$\Delta g^{(3+)}_\text{QED-me,L}$
              &   0.0105         &  -0.0003         &   0.0003         &  -0.0004         &   0.0010     \\
$\Delta g^{(3+)}_\text{QED-me,H}$
              &   0.0000(12)     &   0.0000(4)      &   0.0000         &   0.0000(7)      &   0.0000(6)  \\
Total         &2321.6779(13)     &2321.6607(18)     &2321.6600(17)     &2321.6604(17)     &2321.6593(21) \\ 
\hline
\hline 
\end{tabular}\\
$^a$ Yerokhin {\it et al.} (2020) \cite{yerokhin:2020:022815}.
\end{center}
\end{table*}
The QED corrections are represented by the self-energy and vacuum polarization diagrams in the presence of an external magnetic field.
The computation of the first-order diagrams is well-established and reported, e.g., in Refs.~\cite{blundell:1997:1857, persson:1997:R2499, beier:2000:79, yerokhin:2002:143001, yerokhin:2004:052503, glazov:2004:062104, lee:2005:052501, glazov:2006:330, yerokhin:2017:060501, cakir:2020:062513}.
The second-order contribution $\Delta g^{(2)}_\text{QED}$ comprises the one-electron part $\Delta g^{(2)}_\text{QED-1e}$ (two-loop QED) and the many-electron part $\Delta g^{(2)}_\text{QED-me}$ (screened QED). 
The formal expressions for the screened QED contributions can be found in Refs.~\cite{volotka:2009:033005, glazov:2010:062112, andreev:2012:022510, volotka:2014:253004} 
%
Here, we improve the accuracy of the $\Delta g^{(2)}_\text{QED-me}$ by refining the numerical procedure, see Supplementary Material for details.
%
%
\\
\indent
So far, all the remaining contributions are known only within some approximation.
The one-electron terms, $\Delta g^{(2)}_\text{QED-1e}$ and $\Delta g^{(3+)}_\text{QED-1e}$, are evaluated within the $\alpha Z$ expansion~\cite{aoyama:2019:28, pachucki:2004:150401, pachucki:2005:022108, jentschura:2009:044501, yerokhin:2013:042502, czarnecki:2016:060501, czarnecki:2018:043203, czarnecki:2020:050801}, and the higher-order many-electron QED correction $\Delta g^{(3+)}_\text{QED-me,L}$ is calculated within the recursive perturbation theory~\cite{glazov:2017:46,glazov:2019:173001}, see Supplementary Material for details.
\\
\indent
%
%
In Table~\ref{T3_var_pot}, the results for the QED corrections are presented, obtained with all five binding potentials for both Si$^{11+}$ and Ca$^{17+}$ ions.
For the Coulomb potential, we rely on the results of Refs.~\cite{yerokhin:2017:062511, yerokhin:2020:022815}, these values are given in the second column.
In the third to sixth columns, we present the results obtained in this work.
As seen from this Table, our total values obtained with different screening potentials are close to each other and overlap within their uncertainties.
The total uncertainty is determined by the numerical error in $\Delta g^{(2)}_\text{QED-me}$ and by the estimation of the higher-order effects $\Delta g^{(3+)}_\text{QED-me,H}$.
The latter is assessed as the largest value out of three possible estimations, see the discussion above concerning the $\Delta g^{(3+)}_\text{int,H}$.
We note that in Refs.~\cite{yerokhin:2017:062511, yerokhin:2020:022815}, the uncertainty was estimated similarly, and still there is a significant discrepancy between the Coulomb result and the total results for all the screening potentials.
The reason for this discrepancy can be the calculation of the term $\Delta g^{(2)}_\text{QED-me}$.
As one can see from the Table~\ref{T3_var_pot}, its contribution for the Coulomb potential is much larger than for the screening potentials, and a relatively small shift in its value could lead to an agreement between the results.
\\
\indent
Finally, we average over all the screening potentials and report our final values for the QED correction: $2320.2857(17)\times 10^{-6}$ for silicon and $2321.6601(17)\times 10^{-6}$ for calcium.
\\
\indent
%
%
%
\textit{Total results and conclusion.}
\begin{table}
\caption{\label{T3_g_tot_Si_Ca} Theoretical contributions to the ground-state \emph{g} factor of Li-like $^{28}$Si$^{11+}$ and $^{40}$Ca$^{17+}$ ions. The total theoretical results are compared with the experimental values. The parenthesized numbers indicate the uncertainty of the last digit(s). If no uncertainty is given, all digits are significant.}
\vspace{0.25cm}
\setlength{\tabcolsep}{5pt}
\begin{tabular}{lr@{}lr@{}l} 
\hline
\hline
   Effects     & \multicolumn{2}{c}{$^{ 28}$Si$^{11+}$}  & \multicolumn{2}{c}{$^{40}$Ca$^{17+}$}\\
\hline

Dirac value                      & 1.&998\,254\,753\,3            & 1.&996\,426\,025\,3                    \\
e-e interaction                  & 0.&000\,314\,809\,8(22)        & 0.&000\,454\,291\,0(24)                  \\
QED                              & 0.&002\,320\,285\,7(17)        & 0.&002\,321\,660\,1(17)                  \\
Nuclear recoil                   & 0.&000\,000\,043\,6            & 0.&000\,000\,066\,2                      \\[1mm]
Total theory                     & 2.&000\,889\,892\,4(28)        & 1.&999\,202\,042\,6(29)                  \\
                                 & 2.&000\,889\,893\,7(17)$^a$    & 1.&999\,202\,052\,9(27)$^a$              \\ 
                                 & 2.&000\,889\,896\,3(15)$^b$    & 1.&999\,202\,042(13)$^d$                                       \\
                                 & 2.&000\,889\,894\,4(34)$^c$                                           \\[1mm]
Experiment                       & 2.&000\,889\,888\,45(14)$^c$   & 1.&999\,202\,040\,5(11)$^d$              \\
                                 & 2.&000\,889\,888\,4(19)$^e$                                           \\
\hline\hline
\end{tabular}
\\
\vspace{0.25cm}
$^a$ Yerokhin {\it et al.} (2021) \cite{yerokhin:2021:022814};
$^b$ Yerokhin {\it et al.} (2020) \cite{yerokhin:2020:022815};\\
$^c$ Glazov {\it et al.} (2019) \cite{glazov:2019:173001};
$^d$ K\"ohler {\it et al.} (2016) \cite{koehler:2016:10246};\\
$^e$ Wagner {\it et al.} (2013) \cite{wagner:2013:033003}.
\end{table}
\begin{figure}
    \centering
    \includegraphics[width=0.5\textwidth]{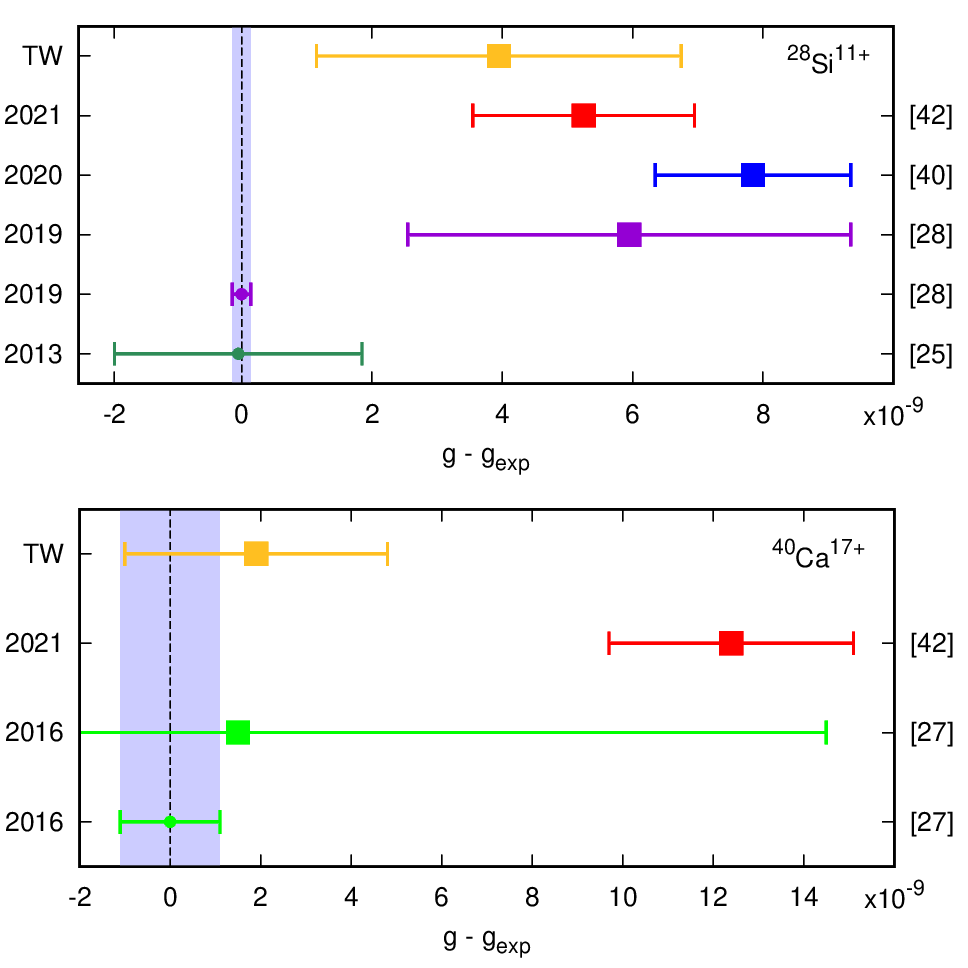}
    \caption{Comparison of the theoretical (squares) and experimental (circles) \emph{g}-factor values for Li-like silicon and calcium obtained in this work (TW) and published previously.}
    \label{fig:comparison}
\end{figure}
In Table~\ref{T3_g_tot_Si_Ca}, we summarize all the theoretical contributions to the \emph{g} factor of Li‐like $^{28}$Si$^{11+}$ and $^{40}$Ca$^{17+}$ ions and compare these results with the previously reported theoretical and experimental data. 
In addition to the corrections $\dgint$ and $\dgqed$ evaluated in this Letter, we also use the nuclear recoil contributions from Refs.~\cite{shabaev:2017:263001,shabaev:2018:032512}. The nuclear polarization effect is negligible for the ions under consideration.
As seen from Table~\ref{T3_g_tot_Si_Ca}, the total uncertainty is still determined by the interelectronic-interaction and QED corrections.
In Fig.~\ref{fig:comparison}, the present and previously published theoretical and experimental results from Refs.~\cite{wagner:2013:033003, koehler:2016:10246, glazov:2019:173001, yerokhin:2020:022815, yerokhin:2021:022814} are depicted together.
In comparison with our previous work~\cite{glazov:2019:173001}, the result for silicon is more accurate, and it is closer to the experimental value.
Comparison with Yerokhin {\it et al.} \cite{yerokhin:2021:022814} shows that for $^{28}$Si$^{11+}$, the results agree within the given uncertainty, while for $^{40}$Ca$^{17+}$, there is a discrepancy of $2.6\sigma$.
We should underline that the individual contributions, $\dgint$ and $\dgqed$, disagree even stronger.
However, these differences partially cancel out each other.
The results of Yerokhin {\it et al.} \cite{yerokhin:2021:022814} differ from the experimental values by $3.1\sigma$ for silicon and by $4.2\sigma$ for calcium.
Meanwhile, our results are much closer to the measurements: $1.4\sigma$ and $0.6\sigma$ deviation, respectively.
We believe that the deviations found in Ref.~\cite{yerokhin:2021:022814} are due to the underestimated uncertainty of the interelectronic-interaction contribution and a possible issue with their calculation of the screened QED term.
\\
\indent
To further improve the accuracy of the total theoretical value of the \emph{g} factor, two-loop many-electron diagrams are to be rigorously evaluated, namely, the three-photon exchange and the two-photon exchange with a self-energy loop.
%
\\
\indent
%
%
\begin{acknowledgments}
We thank V. A. Yerokhin for valuable discussions. This work was supported by DFG (VO1707/1-3) and by RFBR (Grant No. 19-02-00974). A.V.V. acknowledges the financial support by the Government of the Russian Federation through the ITMO Fellowship and Professorship Program. D.A.G. acknowledges the support by the Foundation for the Advancement of Theoretical Physics and Mathematics ``BASIS.'' The computations were performed at the Friedrich Schiller University Jena and supported in part by DFG grants INST 275/334-1 FUGG and INST 275/363-1 FUGG.
\end{acknowledgments}
%
%

%
%
\end{document}